\documentclass[fleqn,12pt,twoside]{article}
\usepackage{espcrc1}

\usepackage{graphicx}
\usepackage{epsfig}
\usepackage[figuresright]{rotating}

\usepackage{amssymb}

\usepackage{amsmath}

\title{Chiral extrapolations of nucleon properties from Lattice QCD\thanks{Talk presented by W. Weise at PANIC 02, Osaka. Work supported in part by BMBF and DFG. Preprint No. TUM-T39-02-23.}}                                                                                                                                                                                                                                                       

\author{T.R. Hemmert\address[TUM]{Physik-Department, Technische Universit\"at M\"unchen, 85747 Garching, Germany},
        M. Procura\addressmark[TUM]\address[ECT]{ECT*, Villa Tambosi, 38050 Villazzano (Trento), Italy} and
        W. Weise\addressmark[TUM]\addressmark[ECT]}

\begin{document}
%
%%%%%%%%%%%%%%%%%%%%%%%%%%%%%%%%%%%%%%%%%%%%%
%
% typeset front matter
\maketitle

The chiral symmetry of QCD is spontaneously broken at low energies, leading to the appearance of Goldstone Bosons. For 2-flavor QCD we identify the resulting three Goldstone Bosons with the physical pion states, the lowest lying modes in the hadron spectrum. In addition to being spontaneously broken, chiral symmetry is broken explicitly by the non-zero quark mass $\hat{m}$ in the QCD Lagrangian. This explicit breaking is responsible for the non-zero pion mass. In the large-condensate scenario with parameter $B_0$ one obtains
\begin{eqnarray}
m_\pi^2=2 \hat{m}B_0\{1+{\cal O}(\hat{m}B_0)\}
\end{eqnarray}
for this connection (now well established for two-flavor QCD). At low energies QCD is represented by a chiral effective field theory ($\chi$EFT) with the dynamics governed by the Goldstone Bosons, coupling to matter fields and external sources. The important aspect for this work is the fact that this $\chi$EFT incorporates {\emph{both}} the information on the spontaneous {\emph{and}} on the explicit breaking of chiral symmetry.
 
We report on recent work utilizing $\chi$EFT to study the quark mass (pion mass) dependence of the magnetic moments and the axial-vector coupling constant of the nucleon. The aim is to explore the feasibility of chiral effective field theory methods for extrapolations of lattice QCD results, so far determined at relatively large quark masses corresponding to pion masses $m_\pi\gtrsim 0.6\,$ GeV, down to physical values of $m_\pi$.
We compare two versions of $\chi$EFT: heavy-baryon chiral perturbation theory in its truncated form restricted to pion and nucleon degrees of freedom only, and an alternative approach which incorporates the $\Delta$(1232), with the $\Delta$-$N$ mass difference (denoted by $\Delta$) treated as a small scale. It turns out that in order to approach the physical values of magnetic moments and $g_A$, the explicit $\Delta$ degree of freedom is crucial.

Our approach includes the leading order $N\Delta$ transition Lagrangian 
\begin{eqnarray}
{\cal L}^{(1)}_{N\Delta}&=&\bar{T}^\mu_i\,\left[c_A w_{\mu}^i\,+\,c_V i {f_{\mu \nu}^{+}}^i S^{\nu}\right]\;N_v\,+\,h.c.\; 
\end{eqnarray}
in consistent power counting schemes described as in ref.\cite{Panic1}. Here $c_V$ and $c_A$ are the vector and axial-vector coupling strengths for the $N\leftrightarrow \Delta$ transition (see ref.\cite{Panic2} for further details).

\begin{figure}[!htb]
  \begin{center}
    \includegraphics*[width=0.65\textwidth,height=13.5cm]{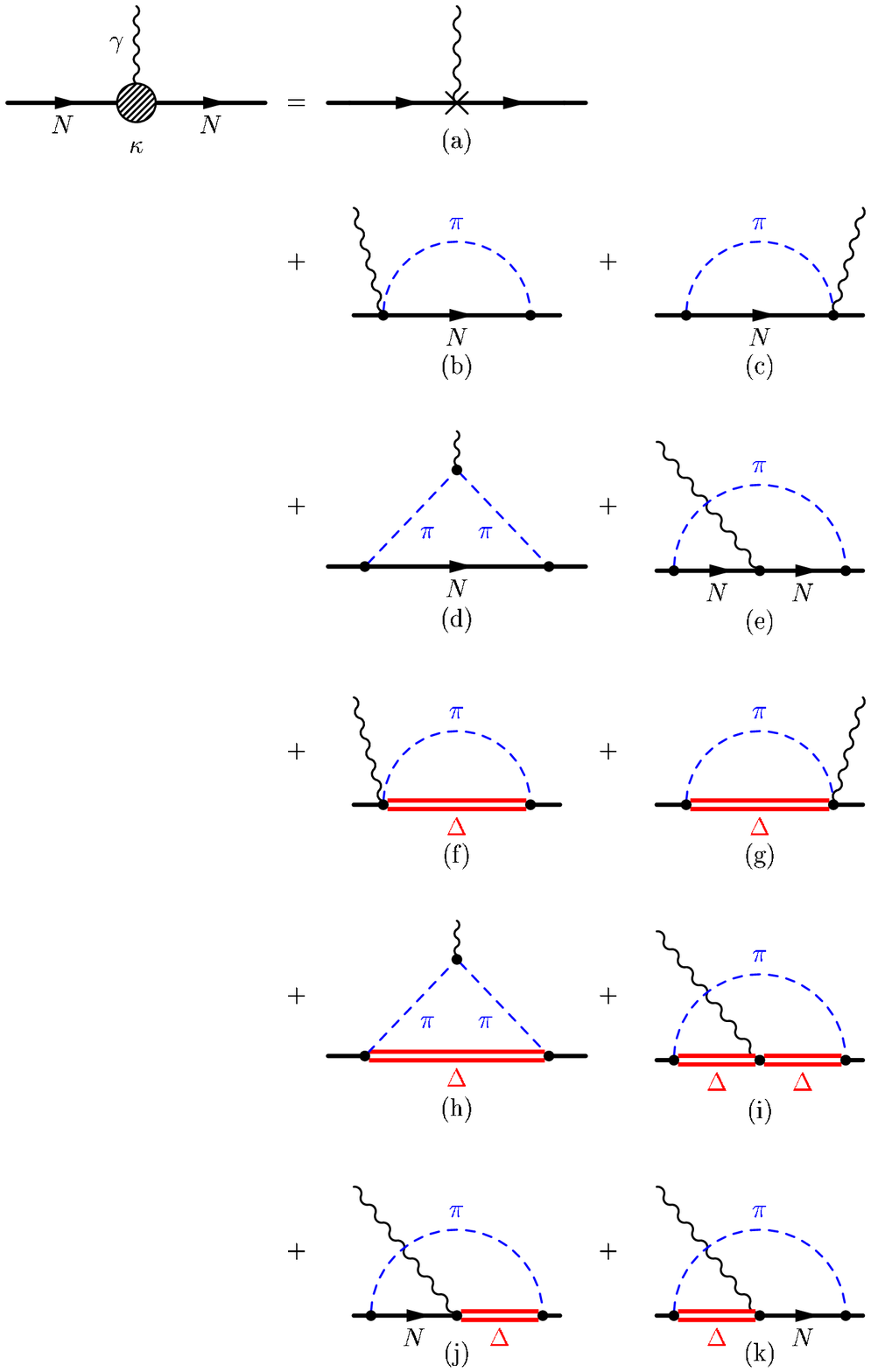}
     \caption{Leading-one-loop diagrams included in the chiral effective field theory approach to nucleon magnetic moments. Diagrams (a)-(e) are the ones incorporated in the truncated heavy-baryon chiral perturbation theory. Diagrams (f)-(k) involve explicit $\Delta$(1232) degrees of freedom.}
\label{Feyndiags}
  \end{center}
\end{figure}

For the calculation of the isovector magnetic moment of the nucleon to one-loop order, the relevant set of diagrams is shown in Fig.\ref{Feyndiags}, with the electromagnetic field treated as an external vector source.

In the corresponding calculation of $g_A$, the external field is replaced by an axial-vector source. Diagrams (d), (f), (g) and (h) of Fig.\ref{Feyndiags} do not exist in this case; instead there is one extra graph involving an axial $\pi\pi N N$ four-point vertex with a pion loop.\\
Consider first the anomalous isovector magnetic moment, $\kappa_V\,=\,\mu_p-\mu_n-1$. The one-loop calculation illustrated in Fig.\ref{Feyndiags} leads to a closed expression of the form
\begin{eqnarray}
\kappa_V=\kappa_V^0-\frac{g_A^2M}{4\pi f_\pi^2}m_\pi+F(m_\pi,\Delta),
\end{eqnarray}
where $\kappa_V^0$ is the ``bare'' moment taken in the chiral limit, $m_\pi\to 0$, with the pion cloud dressing turned off. The explicit expression for the function $F$ is given in ref.\cite{Panic2}. It includes powers and logarithms of the dimensionless variable $m_\pi / \Delta$, with non-analytic quark mass dependence.\\
The small isoscalar anomalous magnetic moment taken to the same leading-one-loop order is analytic in the quark mass: $\kappa_S=\kappa_S^0-const\cdot m_\pi^2$, where the constant parametrizes short-distance physics.

\begin{figure}[!htb]
  \begin{center}
    \includegraphics*[width=0.7\textwidth, height=8cm]{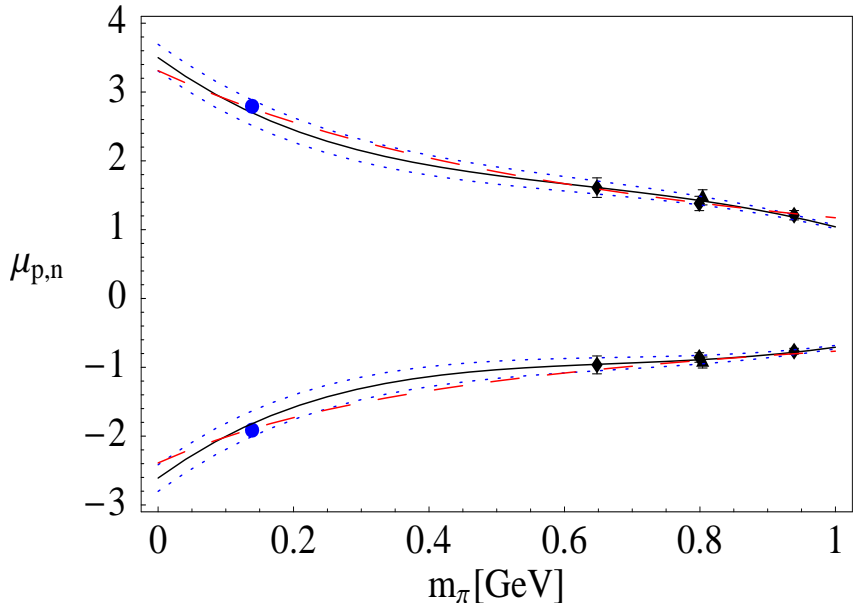}
    \caption{Pion (quark) mass dependence of proton and neutron magnetic moments as computed in ref.\cite{Panic2} (solid), compared to lattice data and a Pad\'e fit formula (dashed) reported in ref.\cite{Panic3}. The empirical $\mu_p=2.793$ and $\mu_n=-1.913$ are also shown.}\label{fit_1}
  \end{center}
\end{figure}

Combining isovector and isoscalar results one obtains the quark (pion) mass dependence of the proton and neutron magnetic moments \cite{Panic2} as shown by the full lines in Fig.\ref{fit_1}. Three parameters ($\kappa_V^0$, the leading $N$$\Delta$ vector coupling $c_V$ and a counterterm representing short-distance dynamics) have been fixed to reproduce quenched lattice results reported in ref.\cite{Panic3}. Although these lattice data correspond to pion masses larger than $0.6$ GeV, one achieves an extrapolation to the physical region which comes remarkably close to the empirical magnetic moments - a non-trivial result. Within the error band given, this result turns out also to be surprisingly close to the Pad\'e fit extrapolation of ref.\cite{Panic3}.

\begin{figure}[!htb]
  \begin{center}
    \includegraphics*[width=0.65\textwidth,height=7cm]{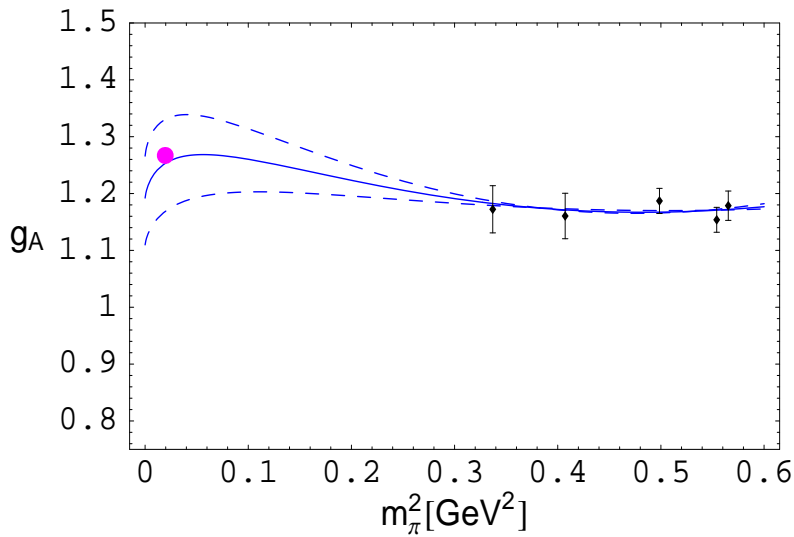}
    \caption{Chiral extrapolation of $g_A$\cite{Panic4} from lattice data\cite{Panic6}. The dashed curves give an impression of the uncertainty range resulting from the input obtained from the $\pi N \to \pi \pi N$ analysis. The empirical value $g_A=1.267$ is also shown.}\label{fit_2}
  \end{center}
\end{figure}

Next, we turn to the axial-vector coupling constant $g_A$. The complete one-loop expression incorporating nucleon and $\Delta$(1232) has the form
\begin{eqnarray}
g_A=g_A^0-\frac{{g_A^0}^3}{(4\pi f_\pi)^2}m_\pi^2+G(m_\pi,\Delta),
\end{eqnarray}
where $g_A^0$ is the axial-vector coupling constant in the chiral limit. The function $G(m_\pi,\Delta)$, specified in ref.\cite{Panic4}, is non-analytic in the quark mass and includes logarithms in $m_\pi/ \Delta$.\\
Results for the evolution of $g_A$ with $m_\pi^2$ (or the quark mass) are shown in Fig.\ref{fit_2}. A counterterm entering in $G$ can be constrained by the chiral perturbation theory analysis of $\pi N\to\pi\pi N$ data \cite{Panic5}. Determining $g_A^0$ and the corresponding axial-$\Delta$-$\Delta$ coupling from the (quenched) lattice data of the QCDSF-UKQCD collaboration \cite{Panic6}, the extrapolation down to the physical region works astonishingly well. The contributions involving the $\Delta$(1232), contained in the function $G$, are crucial in achieving this result. Truncated chiral perturbation theory, with $\pi$ and $N$ degrees of freedom only, fails very badly at this point.

The importance of the explicit $\Delta$(1232) degrees of freedom in the discussion of $g_A$ does not come as a surprise. Recall the Adler-Weisberger sum rule \cite{Panic7},
\begin{eqnarray}
g_A^2&=&1+\frac{2f_\pi^2}{\pi}\int_{m_\pi}^{\infty}\frac{d\omega}{\sqrt{\omega^2-m_\pi^2}}[\sigma_{\pi^+ p}(\omega)-\sigma_{\pi^- p}(\omega)]+{\cal{O}}\left(\frac{m_\pi^2}{M_N^2}\right)\;.
\end{eqnarray}
It relates the surplus of $g_A$ beyond its trivial value $g_A=1$ (for a pointlike, structureless nucleon) to the excess of the $\pi^{+}p$ over the $\pi^{-}p$ cross section, a feature dominated by $\Delta$(1232) resonance excitation.

%
%%%%%%%%%%%%%%%%%%%%%%%%%%%

%%%%%%%%%%%%%%%%%%%%%

\end{document}